\documentclass[11pt]{article}
\usepackage{epsfig}

\begin{document}
%%%%%%%%%%%%%%%%%%%%%%%%%%%%%%%%%%%%%%%%%%%%%%%%%%%%%%%%%%%%%%%%%%%%%%%%%%%
\thispagestyle{empty}
\vspace*{1mm}
\begin{flushright}
MIT-CTP-3008 \\
HLRZ2000\_13 \\
\end{flushright}
\begin{center}
\vspace*{5mm}
{\LARGE Approximate 
Ginsparg-Wilson fermions: 
\vskip2mm
A first test}
\vskip14mm
{\bf Christof Gattringer$\;^{a \dagger}$, Ivan Hip$\;^b$ and C.~B. Lang$\;^c$}
\vskip2mm
$^a\;$Massachusetts Institute of Technology, Center for Theoretical Physics \\
77 Massachusetts Avenue, Cambridge MA 02139 USA
\vskip2mm
$^b\;$NIC - John von Neumann Institute for Computing \\
FZ-J\"ulich, 52425 J\"ulich, Germany
\vskip2mm
$^c\;$Institut f\"ur theoretische Physik, Karl-Franzens Universit\"at Graz\\
Universit\"atsplatz 5, A-8010 Graz, Austria
\vskip17mm
\begin{abstract}
We construct a 4-d lattice Dirac operator $D$ using a  systematical expansion 
in terms of simple operators on the lattice. The Ginsparg-Wilson equation turns
into a system of coupled equations for the expansion  coefficients of $D$. We
solve these  equations for a finite parametrization of $D$ and find  an
approximate solution of the Ginsparg-Wilson  equation. We analyze the spectral
properties of our $D$ for various ensembles of quenched SU(3) configurations. 
Improving  the gauge field action considerably improves the spectral properties
of our $D$.
\end{abstract}
\end{center}
\vskip5mm \nopagebreak \begin{flushleft} \rule{2 in}{0.03cm}
\\ {\footnotesize \ 
${}^\dagger$ Supported by the Austrian Academy of Sciences (APART 654).}
\end{flushleft}
\newpage

\setcounter{equation}{0}
\setcounter{page}{1}
\section{Introduction}
Chiral symmetry in fermionic field theories on the lattice may be implemented
by adding a non-vanishing and in the continuum limit irrelevant right hand side
of the usual anti-commutator of the Dirac operator $D$ with $\gamma_5$. In its
simplest form the resulting so-called Ginsparg-Wilson equation  \cite{GiWi82}
reads (we set the lattice spacing to 1)
\begin{equation}
D \gamma_5 \; + \; \gamma_5 D \; = \; D \gamma_5 D \; .
\label{giwi}
\end{equation}
Based on (\ref{giwi}) chirally symmetric  fermions can be constructed on the
lattice (for reviews of recent developments see 
e.g.~\cite{reviews1,reviews2,reviews3}).

Currently two types of solutions for (\ref{giwi}) are known.  Neuberger
\cite{Neub1} gave an explicit construction, the so-called overlap  operator
based on earlier work, the overlap approach to chiral fermions  on the lattice
\cite{overlap}. The overlap  operator is given by
\begin{equation}
D \; = \; 1 \; - \; A \; \Big( A^\dagger A \Big)^{-\frac{1}{2}} \; \; \; \; \; 
\mbox{with} \; \; \; \; \; A \; = \; 1 \; + \; s \; - \; D_0 \; .
\label{overlapd}
\end{equation}
Here $D_0$ is some decent lattice Dirac operator  which is free of doublers.
Typically the Wilson-Dirac operator is used here but also other choices give
rise to  solutions of (\ref{giwi}); $s$ is a real number which can, at the cost
of  an additional renormalization, be used to optimize the locality properties
of $D$ \cite{overloc}.  The main challenge when using the overlap operator is
the computation of the square root in (\ref{overlapd}),  making simulations
with the overlap operator considerably more expensive than  simulations with
the Wilson operator.

The second solution to (\ref{giwi}) is provided by the fixed point  Dirac
operator \cite{fixpd1,fixpd2} which goes back to the perfect action approach
to  lattice field theories \cite{perfect}. So far, the fixed point Dirac
operator has  been computed only in two dimensions \cite{fix2d,farch,fix2db}
and it has  become clear that its construction  in 4-d is quite a challenging
enterprise.

Recently \cite{Ga00} a new line of attack for solving the  Ginsparg-Wilson
equation has been proposed. The basic idea  is to systematically expand the
most general lattice  Dirac operator $D$ in a series of simple basis operators
on the  lattice. This expanded $D$ is then inserted into the Ginsparg-Wilson
equation. The product $D \gamma_5 D$  on the right hand side of (\ref{giwi})
can be evaluated using techniques from  the analytical hopping expansion. The
result is an expansion of both sides of  (\ref{giwi}); comparing the factors in
front of the contributions to this expansion one finds that the
Ginsparg-Wilson  equation is mapped to an equivalent system of coupled
quadratic equations for the expansion coefficients of $D$. When using finitely
many terms in the expansion of $D$, the system of equations for the expansion
coefficients  can be solved numerically and the corresponding $D$ is an
approximate solution of the Ginsparg-Wilson equation. By including more terms
in the expansion  of $D$ the violation of (\ref{giwi}) can be made arbitrarily
small. 

Here we present first results in 4-d for the discussed method. We  explicitly
construct an approximate solution $D$ of (\ref{giwi})  and study its properties
in different ensembles of quenched SU(3) gauge  configurations.  We find that
already with relatively few terms we obtain a good  approximate solution for
the Ginsparg-Wilson equation.  We discuss properties of the spectrum of our
Dirac  operator and also analyze the role of improving the gauge field action.
We demonstrate that improving the gauge field action considerably improves  the
spectral properties of our $D$. 

We would like to point out three aspects of our approximate solutions of the
Ginsparg-Wilson equation: 
\begin{itemize}
\item In an earlier study \cite{GaHi00} in two dimensions we had  constructed
an approximate solution of (\ref{giwi}) using the outlined ideas.  It turned 
out that essential features of chiral symmetry in the lattice Schwinger  model
were properly described by  our approximate solution of the Ginsparg-Wilson
equation. It will be  interesting to see if our $D$ is able to capture chiral
symmetry also in four dimensions.
\item Our study should help to shed light on the importance of different  terms
in solutions $D$ of the Ginsparg-Wilson equation. In particular for the perfect
action program an economical but still rich enough parametrization of $D$ is
essential \cite{Haetal00}. 
\item Another application of our $D$ is its use as a starting operator $D_0$ in
Neuberger's construction. Using an approximate solution of (\ref{giwi}) in the
projection (\ref{overlapd}) has been proposed  previously \cite{Biholz1} and it
is believed  that an improved $D_0$  speeds up the evaluation of the square
root and helps to overcome the problems with the potential singularity of
(\ref{overlapd}) for small eigenvalues of $A$. 
\end{itemize}

The article is organized as follows: In Section 2 we briefly discuss the 
expansion of $D$ and derive the system of equations equivalent to the 
Ginsparg-Wilson equation. We discuss in Section 3 the boundary conditions and
find a solution of the system of coupled equations. In Section 4 we analyze the
spectral properties of our approximate solution D for ensembles of quenched
gauge field configurations. We start with the free case in Section 4.1 followed
by gauge field ensembles generated  with the standard Wilson gauge action
(Section 4.2). Finally in Section 4.3 we investigate how improving the gauge
field action can further improve the  chiral properties of the lattice Dirac
operator. The article closes with  a discussion in Section 5. In particular we
will come back to  the above mentioned three possible applications of our Dirac
operator.

\setcounter{equation}{0}
\section{Constructing approximate solutions of the \\
Ginsparg-Wilson equation}

To make the article self contained we briefly repeat the basic idea of our 
construction presented in \cite{Ga00}. In a first step we discuss the 
systematic expansion of the most general Euclidean Dirac  operator in a series
of simple operators on the lattice. Subsequently we derive the system of 
quadratic equations for the expansion coefficients  which is equivalent to the
Ginsparg-Wilson equation.

\subsection{Expansion of the most general $D$} 

Typically the derivative term on the lattice is discretized by the following
nearest neighbor term (we set the lattice spacing to 1):
\begin{equation}
\frac{1}{2} \sum_{\mu = 1}^4 \gamma_\mu \left[
U_\mu(x) \delta_{x+\hat{\mu},y} - 
U_\mu(x - \hat{\mu})^{-1} \delta_{x-\hat{\mu},y} \right] \; .
\label{example1}
\end{equation}
However, it is perfectly compatible with all the symmetries to 
instead discretize the derivative term using
\begin{equation}
\frac{1}{4} \sum_{\mu = 1}^4 \gamma_\mu \left[
U_\mu(x) U_\mu(x\!+\!\hat{\mu}) \; \delta_{x+2 \hat{\mu},y} \; - 
\; U_\mu(x\!-\!\hat{\mu})^{-1} U_\mu(x-2\hat{\mu})^{-1} \;
\delta_{x\!-\!2\hat{\mu},y} \right] ,
\label{example2}
\end{equation}
and there are many more terms one could think of. Thus an ansatz for  the most
general Dirac operator $D$ must allow for a superposition of all of the
possible discretizations for the derivative term. 

The terms in  (\ref{example1}) and (\ref{example2}) can be characterized by
simple  paths on the lattice. The first example (\ref{example1}) consists  of a
single hop in positive $\mu$-direction with a plus sign and a single hop in
negative direction with a minus sign. Similarly the  second example consists of
two hops in positive (negative) $\mu$-direction.  We now introduce a shorthand
notation for such paths using an ordered list of the directions of the links of
the path.  We denote a path of length $n$ on the lattice by 
\begin{equation}
<l_1,l_2, ... \; l_n>, 
\label{notation}
\end{equation}
with the $l_i$ giving the directions of the subsequent hops,  i.e.~$l_i \in \{
\pm1,\pm2,\pm3$, $\pm4\}$. It is implicitly understood, that each link is
dressed with the corresponding link variable $U_\mu(x)$. Using this notation
the terms from the two examples (\ref{example1}) and (\ref{example2}) are
denoted as
\begin{equation}
\frac{1}{2} \sum_{\mu} \gamma_\mu \sum_{l = \pm \mu} s(l) \;
 < l > \; ,
\end{equation}
and
\begin{equation} 
\frac{1}{4} \sum_{\mu} \gamma_\mu \sum_{l = \pm \mu} s(l) \;
 < l, l > \; .
\end{equation}
We use the abbreviation $s(l)$ for sign$(l)$. Due to translation invariance the
form of the derivative terms, and thus of the paths,  is the same at all
lattice points such that in our notation no reference to the starting point for
the paths is necessary.

In order to  remove the doublers one also needs a term which in momentum space
can  distinguish between $p_\mu = 0$ (physical modes) and $p_\mu = \pi$
(doublers). Such a term is provided by the standard Wilson term. Due to the
symmetries this term has to come with  1\hspace*{-1.0mm}I in spinor space.
Again we will allow for all possible  terms. We generalize our $D$ further, by 
including all terms also for the remaining elements  $\Gamma_\alpha$ of the
Clifford algebra, i.e.~tensors, pseudovectors and  the pseudoscalar. Thus the
emerging lattice Dirac operator has the following form:
\begin{equation}
D \; = \; \sum_{\alpha = 1}^{16} \Gamma_\alpha \; 
\sum_{p \in {\cal P}^\alpha} c_p^\alpha \; 
<l_1,l_2, ... \; l_{|p|}> .
\label{ansatz}
\end{equation}
To each generator  $\Gamma_\alpha$ of the Clifford algebra we assign a set
${\cal P}^\alpha$ of paths $p$, each $p$ given by some  ordered set of links
$<l_1,l_2, ... l_{|p|}>$ (compare (\ref{notation})) where $|p|$ denotes the
length of the path $p$. Each path is weighted with some complex weight
$c_p^\alpha$.

The next step is to impose on $D$ the symmetries which we want to  maintain:
Translation and rotation invariance and invariance under  C and P. In addition
we require our $D$ to be $\gamma_5$-hermitian,  i.e.~we require $\gamma_5 D
\gamma_5 = D^\dagger$. This property can  be seen to correspond to what leads
to the CPT theorem in Minkowski  space, i.e.~the vector generators $\gamma_\mu$
come with a derivative  term etc. 

Translation invariance has already been briefly mentioned  above and requires
the sets   ${\cal P}^\alpha$ of paths and their coefficients  to be independent
of the starting point.  Rotation invariance implies that a path and its rotated
image have the same weight. Parity implies that for each path $p$ (with
coefficient $c_p^\alpha$) we must include the parity-reflected copy with
coefficient $s_{parity}^\alpha \cdot c_p^\alpha$ where the signs
$s_{parity}^\alpha$ are  defined by $\gamma_4 \Gamma_\alpha \gamma_4 =
s_{parity}^\alpha  \cdot \Gamma_\alpha$. 

Of importance are the symmetries C and  $\gamma_5$-hermiticity. It is easy to
see that both of them imply a relation between the coefficient for a path $p$
and the coefficient of the inverse path $p^{-1}$. Implementing both these
symmetries restricts all  coefficients $c_p^\alpha$ to be either real or purely
imaginary. Furthermore we find that the coefficient for a path $p$ and the
coefficient for its inverse $p^{-1}$ are equal up to a sign $s_{charge}^\alpha$
defined  by $C \Gamma_\alpha C = s_{charge}^\alpha \cdot  \Gamma_\alpha^T$,
where $T$  denotes transposition and $C$ is the charge conjugation matrix.

When implementing all these symmetries we find that paths in our ansatz  become
grouped together where -- up to sign factors -- all paths in a group come with
the same coefficient. We can now write down our most  general Dirac operator on
the lattice in the form (compare \cite{Ga00} and see also \cite{Haetal00} for
an equivalent derivation using a slightly different notation):
\begin{eqnarray}
&D& \! \equiv \mbox{1\hspace*{-1.0mm}I} \Big[ s_1 \! <> \; + \;
s_2\! \sum_{l_1} <l_1> 
\; + \; s_3 \!\sum_{l_2 \neq l_1} <l_1,l_2> 
\; + \; s_4 \!\sum_{l_1} <l_1,l_1> \; ... \Big] 
\nonumber
\\ 
&+& \sum_{\mu} \gamma_\mu \sum_{l_1 = \pm \mu} s(l_1) \Big[ \;
v_1\!< l_1 > \; + \; v_2\!\sum_{l_2 \neq \pm \mu} [ <l_1,l_2> + <l_2,l_1> ] 
\nonumber 
\\
& & \hspace*{8cm}
+ \; v_3\!< l_1,l_1> \; ... \; \Big]
\nonumber
\\
&+& \sum_{\mu < \nu} \gamma_\mu \gamma_\nu \sum_{{l_1 = \pm \mu
\atop l_2 = \pm \nu}} s(l_1)\; s(l_2)
\sum_{i,j = 1}^2 \epsilon_{ij} \Big[ \; t_1 <l_i,l_j> \; ... \; \Big]
\nonumber
\\
&+& \!\! \sum_{\mu < \nu < \rho} \gamma_\mu \gamma_\nu \gamma_\rho
\!\! \sum_{{l_1 = \pm \mu, l_2 = \pm \nu \atop l_3 = \pm \rho}} \!\! 
s(l_1)\; s(l_2) \; s(l_3)
\sum_{i,j,k = 1}^3 \epsilon_{ijk} \Big[ \; a_1 <l_i,l_j, l_k> \; ... \; \Big]
\nonumber
\\
&+& \gamma_5 \!\!
\!\! \sum_{{l_1 = \pm 1, l_2 = \pm 2 \atop 
l_3 = \pm 3, l_4 = \pm 4}} \!\! 
s(l_1)\; s(l_2) \; s(l_3) \; s(l_4)
\sum_{i,j,k,n = 1}^4 \epsilon_{ijkn} \Big[ \; p _1 <l_i,l_j, l_k, l_n> 
\; ... \; \Big]\; .
\nonumber \\
\label{dexp}
\end{eqnarray}
By $\epsilon$ we denote  the totally anti-symmetric tensors with 2,3 and 4
indices. We choose the  normalization of the elements of the Clifford algebra
such that the  elements appear as all possible products of the $\gamma_\mu$
without any  extra factors of $i$. For this  normalization the symmetries C and
$\gamma_5$-hermiticity render all coefficients $s_i,v_i,t_i,a_i$ and $p_i$
real. To be specific, we use the  Euclidean chiral representation for the
$\gamma_\mu=\gamma_\mu^\dagger$.  In principle it would be possible to
generalize $D$ further by multiplying each term with a polynomial of traces of 
gauge field variables around closed loops on the lattice, but we do not include
this possibility here.

The above mentioned structure of paths appearing in groups is obvious. The
paths in each group are related by symmetries and up to the sign factors have
to come with the same real coefficient. All paths within a  group have the same
length.

It has to be stressed, that in (\ref{dexp}) for each generator we show  only
the leading terms of an infinite series of groups of paths. The dots indicate
that we omitted groups with paths that are longer than the terms  we display.
It is known \cite{noultraloc} that no ultra-local  solutions of the
Ginsparg-Wilson equation exist and thus an expansion for a solution of
(\ref{giwi}) necessarily contains infinitely many terms. Eventually we will
work with a finite $D$  (the corresponding truncation will be discussed in
Section 3)  but for the moment we keep deriving the method in its most general
form,  i.e.~containing no truncation. 

\subsection{The system of coupled equations corresponding to the \\
Ginsparg-Wilson equation}

Let us now insert our expanded Dirac operator $D$ into the Ginsparg-Wilson 
equation. To that purpose we multiply (\ref{giwi})  with $\gamma_5$ from the
left, bring the terms linear in $D$ to the right-hand side and define:
\begin{equation}
E \; \equiv \; - \; D \; - \; \gamma_5 D \gamma_5 \; + \; 
\gamma_5 D \gamma_5 D \; .
\label{edef}
\end{equation}
We remark that $E$ is hermitian since we implemented $\gamma_5$-hermiticity for
$D$. Finding a solution $D$ of the Ginsparg-Wilson equation corresponds to
having $E = 0$. Evaluating the linear part of $E$ is straightforward: When
evaluating the product $\gamma_5 D \gamma_5$ we find that the terms with an odd
number of $\gamma_\mu$, i.e.~vector- and pseudovector terms pick up a minus
sign, while the other terms remain unchanged. Thus when adding the two linear
terms we find that the terms with an odd  number of $\gamma_\mu$ cancel, while 
the other terms pick up a factor of 2.

The next step is to compute the quadratic term $\gamma_5 D \gamma_5 D$.  Here
we have to multiply the various terms appearing in $D$. Each term  is made out
of two parts, a generator of the Clifford algebra and a group of  paths. The
multiplication of two of these terms proceeds in two steps:  First the two
elements of the Clifford algebra are multiplied giving   again an element of
the algebra. In the second step we have to multiply  the paths of our two
terms. This multiplication can be noted very  conveniently in our notation,
where multiplication of two paths simply  consists of writing the paths into
one long path: 
\begin{equation}
<l_1,l_2...l_n> \times < l_1^\prime,l_2^\prime ... l_{n^\prime}^\prime> 
\; \; = \; \; 
<l_1,l_2...l_n,l_1^\prime,l_2^\prime ... l_{n^\prime}^\prime> \; .
\label{pathprod}
\end{equation} 
It is straightforward to establish this rule by translating back to the 
algebraic expression of our examples (\ref{example1}),(\ref{example2}) and
performing the multiplication in this notation. It can happen that after 
multiplying two paths a hop in some direction $l_i$ is immediately followed by
its inverse $-l_i$. These two hops then cancel each other and we  find 
\begin{equation}
< l_1 ... l_{i-1},l_i,-l_i,l_{i+1} ... l_n> \; \; = \; \; 
< l_1 ... l_{i-1},l_{i+1} ... l_n> \; . 
\label{backtrack}
\end{equation}
This rule is used to reduce all products of  paths appearing in $\gamma_5 D
\gamma_5 D$ to their true length. In a  final step we decompose the product
terms into groups related by the  symmetries in the same way as we did above
when constructing the most  general ansatz for $D$. Adding the linear and
quadratic terms of  (\ref{edef}) we end up with the following expansion for
$E$:
\begin{eqnarray}
&E& \! \equiv \mbox{1\hspace*{-1.0mm}I} \Big[ e^s_1 \! <> \; + \;
e^s_2\! \sum_{l_1} <l_1> 
\; + \; e^s_3 \!\sum_{l_2 \neq l_1} <l_1,l_2> 
\; + \; e^s_4 \!\sum_{l_1} <l_1,l_1> \; ... \Big] 
\nonumber
\\ 
&+& \sum_{\mu} \gamma_\mu \sum_{l_1 = \pm \mu} s(l_1) \Big[ \;
e^v_1\!\sum_{l_2 \neq \pm \mu} [ <l_1,l_2> - <l_2,l_1> ] 
\; ... \; \Big]
\nonumber
\\
&+& \sum_{\mu < \nu} \gamma_\mu \gamma_\nu \sum_{{l_1 = \pm \mu
\atop l_2 = \pm \nu}} s(l_1)\; s(l_2)
\sum_{i,j = 1}^2 \epsilon_{ij} \Big[ \; e^t_1 <l_i,l_j> \; ... \; \Big]
\nonumber
\\
&+& \!\! \sum_{\mu < \nu < \rho} \!\!\!
\gamma_\mu \gamma_\nu \gamma_\rho
\!\!\!\!\!\!\!
\sum_{{l_1 = \pm \mu, l_2 = \pm \nu \atop l_3 = \pm \rho}} \!\!\!\!\! 
s(l_1)\; s(l_2) \; s(l_3)\!\!\!
\sum_{i,j,k = 1}^3 \epsilon_{ijk} \Big[ 
\sum_{l_4 \neq \pm\mu,\nu,\rho}\!\! \Big\{
e^a_1 [<l_i,l_j,l_k,l_4> 
\nonumber \\
&-& <l_4,l_i,l_j,l_k>] + 
e^a_2 [<l_i,l_4,l_j,l_k> - <l_i,l_j,l_4,l_k>] \Big\}  
... \Big]
\nonumber
\\
&+& \gamma_5 \!\!
\!\! \sum_{{l_1 = \pm 1, l_2 = \pm 2 \atop 
l_3 = \pm 3, l_4 = \pm 4}} \!\! 
s(l_1)\; s(l_2) \; s(l_3) \; s(l_4)
\sum_{i,j,k,n = 1}^4 \epsilon_{ijkn} \Big[ \; e^p _1 <l_i,l_j, l_k, l_n> 
\; ... \; \Big]\; .
\nonumber \\
\label{eexp}
\end{eqnarray}
We remark that all the algebraic steps leading to the expansion of $E$ are
straightforward to formalize and for the higher orders in the expansion  we
used a computer program. As for $D$, the expansion of $E$ is an infinite series
and  we display here only the leading groups of paths. The coefficients
$e^\alpha_i$ are now quadratic polynomials in the original coefficients 
$s_i,v_i,t_i,a_1$ and $p_i$ given by 
\begin{eqnarray}
e^s_1 & = & - \; 2s_1 + s_1^2 + 8s_2^2 + 48s_3^2 + 8s_4^2 + 8v_1^2 + 96v_2^2
+ 8v_3^2 + 48t_1^2 + 192a_1^2 
\nonumber \\
& &+ \; 384p_1^2 \; ... \; ,
\nonumber \\
e^s_2 & = & - \; 2s_2 + 2s_1s_2 + 12 s_2 s_3 + 2s_2 s_4 + 12 v_1 v_2 
+ 2v_1 v_3 \; ... \; ,
\nonumber \\
e^s_3 & = & - \; 2 s_3 + 2s_1 s_3 + s_2^2 + 4s_3^2 + 2s_3s_4 + 4v_2^2 + 
2v_2 v_3 \; ... \; ,
\nonumber \\
e^s_4 & = & - \; 2s_4 + 2s_1s_4 + s_2^2 + 6s_3^2 - v_1^2 - 6t_1^2 
- 24a_1^2 - 48 p_1^2 \; ... \;  ,
\nonumber \\
e^v_1 & = & - \; s_2v_1 - 4s_3v_2 - 2s_4v_2 - s_3v_3 - v_3t_1 - 
4v_2t_1 \; ... \; ,
\nonumber \\
e^t_1 & = & - \; 2t_1 + 2s_1t_1 -  2 s_4t_1 - v_1^2 - 4v_2^2 - 2v_2v_3 
- 4t_1^2 + 8 v_1a_1 - 8a_1^2 
\nonumber \\
& & + \; 16t_1p_1 \; ... \; ,
\nonumber \\
e^a_1 & = & - \; s_2 a_1 + v_2 t_1  - v_3 p_1  \; ... \; ,
\nonumber \\
e^a_2 & = & - \; v_2 t_1 \; ... \; ,
\nonumber \\
e^a_3 & = & - \; s_2 a_1  - 2 v_2 p_1  \; ... \; ,
\nonumber \\
e^a_4 & = & - \; 2s_2 a_1 + 2v_2 t_1 - 4v_2 p_1 \; ... \; ,
\nonumber \\
e^p_1 & = & - \; 2p_1 + 2s_1 p_1 - 2s_4p_1 - 2v_1a_1 + t_1^2 \; ... \; .
\label{qsyst}
\end{eqnarray}
Due to different symmetry properties ($E$ is hermitian, $D$ is
$\gamma_5$-hermitian) there are  terms in $E$ which do not occur in $D$ and
vice versa. Each coefficient itself is an infinite series of terms. For a
solution  of the Ginsparg-Wilson equation we must have $E = 0$. It is easy to
see, that the groups of paths appearing in the expansion (\ref{eexp}) are 
linearly independent and hence for $E = 0$ all coefficients $e^\alpha_i$ have
to vanish simultaneously. Thus we have rewritten the problem of  finding a
solution of the Ginsparg-Wilson equation to solving the system  (\ref{qsyst})
of coupled quadratic equations (set all right-hand sides $e^\alpha_i = 0$). 

\setcounter{equation}{0}
\section{Solving the system of coupled equations}

In the last section we have shown that the Ginsparg-Wilson equation is 
equivalent to a system of coupled quadratic equations for the expansion 
coefficients of $D$. In this section we now truncate our expansion of $D$ and
find solutions for the coupled equations. Before we do this  let us discuss the
boundary conditions which have to be added to  the system (\ref{qsyst}).

\subsection{Boundary conditions}

For the free case the situation is simple: In this case  we can compute the
Fourier transform $\hat{D}(p)$ of $D$.  For small momenta the massless Dirac
operator should obey 
\begin{equation}
\hat{D}(p) \; \equiv \; i \not \hspace{-1mm} p \; + \; {\cal O}(p^2) \; .
\label{dfour}
\end{equation}
This leads to two more supplementary equations for the coefficients, one for
the constant term to vanish and the second one  sets the slope of the
dispersion relation equal to one:
\begin{eqnarray}
0 & = & s_1 \; + \; 8\; s_2 \; + \; 48 \; s_3 \; + \; 8 \; s_4 \; ... \; \; ,
\label{freedbc1} \\
1 & = & 2\; v_1 \; + \; 24 \; v_2 \; + \; 4 \; v_3 \; ... \; \; .
\label{freedbc2}
\end{eqnarray}
We remark, that (\ref{freedbc1}) is implicitly guaranteed by the
Ginsparg-Wilson equation, since the Ginsparg-Wilson  circle\footnote{It is well
known, that an exact, $\gamma_5$-hermitian solution  of the Ginsparg-Wilson
equation has its spectrum on the so-called  Ginsparg-Wilson circle, i.e.~the
circle of radius 1 with center 1 in the complex plane.} runs through the
origin. When we construct  approximate solutions below it will be necessary to
explicitly implement  (\ref{freedbc1}). When we started to solve the system of
equations (\ref{qsyst}) respecting  the boundary conditions (\ref{freedbc1}),
(\ref{freedbc2}) we  found, that for this setting one has to include a
relatively  large number of terms in the expansion (\ref{dexp}) of $D$.  Since
every  new term in $D$ drives up the cost of a numerical treatment of our Dirac
operator we decided to allow for an additional freedom in our equations. The
idea is to allow some $\beta$-dependence by introducing a new  parameter,
similar in spirit to the  variable $s$ in the overlap construction
(\ref{overlapd}), which may be used  to improve locality of $D$ , i.e.~to make 
higher terms in the expansion (\ref{dexp})  of $D$ less important.  At the same
time we do not want to change our coupled equations (\ref{qsyst}) since they
guarantee  that -- up to a certain order -- we deal with a solution of the 
Ginsparg-Wilson equation. Thus we only modify the boundary conditions
(\ref{freedbc1}), (\ref{freedbc2}).

Before we discuss the modifications let us first develop the idea for the case
of the standard Wilson action. Wilson's lattice Dirac operator $D_W$ is a sum
of a constant and a hopping term $H$
\begin{eqnarray}
D_W \; & = & 4 \; - \; \kappa \; H \; , 
\nonumber \\
H_{x,y} & = & \sum_\mu \Big\{ [ 1 - \gamma_\mu ] \; U_\mu(x) \;  
\delta_{x+\mu, y} + 
[ 1 + \gamma_\mu ] \; U_\mu(x-\mu)^{-1} \delta_{x-\mu, y} \Big\} \; .
\nonumber \\
\label{dwil}
\end{eqnarray}
For the free case the Fourier transform of the hopping matrix is given by
$\hat{H}(p) = 8 - i \; 2 \not \hspace{-1mm} p +{\cal O}(p^2)$.  Thus the
boundary condition corresponding to (\ref{dfour}) reads
\begin{equation}
4 \; - \; 8\, \kappa  \; + \; 2\,i\,\kappa \not \hspace{-1mm} p 
\; + \; {\cal O}(p^2) \; \equiv \; i \not \hspace{-1mm} p 
\; + \; {\cal O}(p^2) \; ,
\label{freewibc}
\end{equation}
and from this equation one finds that for the free case the correct value  of
the parameter is given by $\kappa = 1/2$.  In fact, in the notation of our
parametrization of $D$ we have $s_1=4$, $s_2= -\,\kappa$ and $v_1=\kappa$ with
all other coefficients  vanishing. The condition (\ref{freewibc}) for the
Wilson operator therefore corresponds to both our boundary  conditions
(\ref{freedbc1}) and (\ref{freedbc2}).  In principle one could generalize the
Wilson action and lift the degeneracy between $-s_2$ and $v_1$ which would
amount to two  boundary conditions similar to (\ref{freedbc1}),
(\ref{freedbc2}) instead of the single condition (\ref{freewibc}).

The situation changes when we include gauge fields. One finds 
that in order to drive the system critical one needs to
change the value of $\kappa$. E.g.~when analyzing quenched SU(3) gauge  theory
at $\beta = 6.0$ one finds $\kappa = 0.624$ (see below). Obviously this value
of kappa is not a solution of the boundary condition (\ref{freewibc})  but
instead solves the more general equation
\begin{equation} 
4 \; - \; 8\, \kappa \, z \;  +  \; 2\,i\,\kappa \, z \, \not \hspace{-1mm} p 
\; + \; {\cal O}(p^2) \; \equiv \; i \not \hspace{-1mm} p 
+  {\cal O}(p^2) \; .
\label{genwibc}
\end{equation}
Here $z$ is a real function of $\beta$. For quenched gauge fields at  $\beta =
6$, we find e.g. $z = 1.603 = 1/0.624$. 

We now generalize the boundary conditions (\ref{freedbc1}), (\ref{freedbc2}) 
analogous to (\ref{genwibc}) being a generalization of (\ref{freewibc}).  In
the boundary condition we multiply each coefficient by a power $z^n$ of
some real, $\beta$-dependent number $z$.  The exponent $n$ is  given by the
number of hops of the corresponding term in our expansion (\ref{dexp}). Thus 
e.g.~the coefficient $s_1$, which is the constant term, does not get changed at
all, $s_2$ which is the coefficient for the single hop obtains a factor of $z$,
$s_3$ which corresponds to two hops becomes multiplied by $z^2$ et cetera.
We allow for two different coefficients $z_s$ and
$z_v$ in the scalar and vector sectors. 
Our generalized boundary conditions then read
\begin{eqnarray}
0 & = & s_1 \; + \; 8\; s_2 \; z_s \; + \; 48 \; s_3 \; z_s^2 \; + 
\; 8 \; s_4 \; z_s^2 \; ... \; ,
\nonumber \\
1 & = & 2\; v_1 \; z_v \; + \; 24 \; v_2 \; z_v^2 \; + \; 4 \;
v_3 \; z_v^2 \; ... \; .
\label{rendbc}
\end{eqnarray}
The two parameters $z_s(\beta)$ and $z_v(\beta)$ can be used to optimize the 
properties of $D$ while at the same time maintaining unchanged the equations
(\ref{qsyst}) which are equivalent to the Ginsparg-Wilson equation. 

We emphasize, that these two supplementary conditions are just a practical
means to effectively reduce the number of necessary terms in the  Dirac
operator. In some sense they serve as a guiding principle to select more local
actions in the huge space of possible actions satisfying the Ginsparg-Wilson
condition.

We can now restrict ourselves to a much smaller set of terms in the expansion
(\ref{dexp}) of $D$. Once this set is chosen, $z_s$ and $z_v$ will be
determined by optimizing the properties of the spectrum of $D$ near the origin.
The parameter $z_s$ is fixed by the requirement that the small  eigenvalues of
$D$ fall on the Ginsparg-Wilson circle (i.e. $m=0$); ~$z_v$ can   e.g.~be fixed
by requiring  the slope of the $\pi$-dispersion relation to be equal to 1. In
this preliminary study we will simply determine $z_v$ by optimizing the
alignment of larger eigenvalues along the circle. 

It is important to remark, that already including only a few terms in the
expansion (\ref{dexp}) of $D$ and solving the corresponding system 
(\ref{qsyst}) considerably orders the small eigenvalues of $D$ (see below). 
Thus no complicated fine-tuning procedure is necessary for $z_s$ and $z_v$ and
they can simply be determined by analyzing the spectrum  of $D$ on a few
background gauge field configurations.

Finally we remark, that another feature which one would like to  implement is
${\cal O}(a)$ improvement \cite{oaimp}.  It is known, that an exact solution of
the  Ginsparg-Wilson equation is already ${\cal O}(a)$ improved. However, here
we will discuss only an approximate solution and it is useful to require 
${\cal O}(a)$  improvement independently of the Ginsparg-Wilson equation.
${\cal O}(a)$  improvement may be achieved by adding to $D$ the so-called
clover-leaf  term $c_{sw} \frac{1}{2} \sigma \cdot F$ \cite{ShWo85}. At tree
level the coefficient  $c_{sw}$ equals the factor in front of the Laplace-type
contribution to $D$.  For non-perturbative improvement $c_{sw}$ can e.g.~be
determined using the Schr\"odinger functional \cite{schrfun}. Since we do not
attempt to  determine $c_{sw}$ non-perturbatively here we quote the condition 
at tree level (i.e.~$c_{sw} = 1$). It is obtained by expanding the terms in 
the tensor sector of our $D$ for small lattice spacing, extracting their 
$\frac{1}{2}\, \sigma \cdot F$ content and setting the factor in front of this
term equal to the factor of the Laplace-term contribution of $D$. One finds
(compare also \cite{Haetal00})
\begin{equation}
s_2 \; + \; 12 \; s_3 \; + \; 4 \; s_4 \; ... \; = \; 
4 \; t_1 \; + \; 32 \; t_2 \; + \; 16 \; t_3 \; ... \; \; .
\label{improve}
\end{equation}
It is straightforward to include $c_{sw}$ as a free parameter in 
(\ref{improve}) but as already remarked above we do not attempt a 
nonperturbative evaluation of $c_{sw}$ here and thus work with the tree level
equation (\ref{improve}) throughout this paper.

At this point we would like to comment on the fate of the doubler modes.  
When analyzing our solutions for the free case we always find that the 
eigenvalues for the doublers (at least one component $p_\mu$ of the
four-momentum equals $\pi$) are located near 2 in the complex plane as is 
the case for exact, doubler-free solutions of the Ginsparg-Wilson equation.   
Thus our approximation scheme
already takes care of the doublers and we do not need to supplement the 
boundary condition (\ref{dfour}) by an additional equation for the doublers.

\subsection{Truncation of $D$ and numerical solution of the system of coupled
equations}

In the last section the set of equations (\ref{qsyst}) together with  the
boundary conditions (\ref{rendbc}) and (\ref{improve}) was discussed. In
principle the system (\ref{qsyst}) contains infinitely many equations,  each of
them with infinitely many terms. Also the boundary conditions contain
infinitely many terms. The next  step thus will be a truncation of the
expansion (\ref{dexp}) for $D$ which  will reduce the infinite problem to a
finite problem. 

\begin{figure}[tb]
\begin{center}
\epsfig{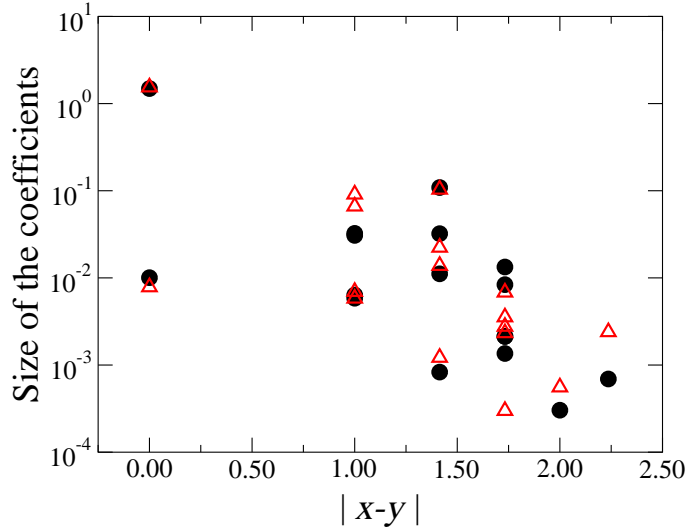}
\caption{Size of the coefficients as a function of the distance $|x-y|$
on the lattice. We use full circles for all the coefficients in the free case, 
the triangles represent 
the coefficients of the quenched ensemble at $\beta = 6.0$ 
(cf. Table \ref{coeffvals}).
\label{coefflog}}
\end{center}
\end{figure}

Any reasonable truncation should have a parameter which controls the size of
the remainder of the approximation. In our case such a parameter is given by
the length of the paths in each term in the expansion (\ref{dexp}) of $D$. We
find that roughly the size of the coefficients  $s^\alpha_i, v^\alpha_i,
t^\alpha_i, a^\alpha_i, p^\alpha_i$  decreases exponentially as the length of
the paths in the  corresponding terms increases (compare Table \ref{coeffvals}
in the appendix). Thus neglecting terms with longer paths  provides a natural
cutoff scheme for the expansion of $D$. The exponential decrease of the
coefficients not only justifies our cutoff, but is also important for the
physics described by $D$.  A solution of the Ginsparg-Wilson equation cannot be
ultra-local   \cite{noultraloc}. However, in order to remain in the  correct
universality class, $D$ has to be local, i.e.~$D_{x,y}$ has to decrease
exponentially  as the distance $|x-y|$ on the lattice increases. In
Fig.~\ref{coefflog} we plot the size of the coefficients as a function of $|x -
y|$  on a logarithmic scale. We show the coefficients for the free case 
(full circles are used to represent all the coefficients)
as well
as those used for the quenched gauge field  ensemble (triangles) 
from the Wilson gauge
action at $\beta = 6.0$ (cf. Table \ref{coeffvals}).  The coefficients for the
other three  ensembles show a similar behavior.  The size of the coefficients
is consistent with being bound by an exponential decay
with $|x-y|$. Thus we have verified, that we
approximate a healthy, local solution of (\ref{giwi}).

Let us now come to the more practical aspects of the truncation. It is clear 
that adding more terms in the expansion (\ref{dexp}) of $D$ improves the
quality of the approximation of a solution of the Ginsparg-Wilson  equation. On
the other hand each new term drives up the numerical cost. Here we present an
operator which manages to give a decent approximation of a solution of
(\ref{giwi})  but at the same time is still relatively cheap to simulate. When
constructing the operator we started with a simple parametrization and added
only terms which gave rise to a considerable improvement  of the spectral
properties. We stopped adding terms when   our $D$ had a satisfactory balance
of good chiral properties and low numerical cost. 

Also the choice of equations  from the set (\ref{qsyst}) allows for some
freedom (the boundary conditions (\ref{rendbc}) and (\ref{improve}) are always
implemented). This freedom  again goes back to the non-existence
\cite{noultraloc} of ultra-local solutions of (\ref{giwi}). If we could solve
all equations (\ref{qsyst}) for a finite parametrization (\ref{dexp}) of $D$
this would, however, amount to an  ultra-local solution. The loophole out of
this dilemma is the fact that the system (\ref{qsyst}) is always
overdetermined. Thus we can only solve the equations corresponding to the
leading terms of the expansion  (\ref{eexp}) of $E$, where again the length of
the paths is the expansion parameter.

The solution which we present here has altogether 17 terms in the expansion
(\ref{dexp}) of $D$. The maximum path length is 4 and we only included terms in
the scalar, vector and tensor sectors. Except for one term ($s_6$) all terms
have paths on the hypercube. For the vector and tensor sectors  we allowed only
terms up to length 3 on the hypercube, since longer terms in these sectors 
create many new entries in the fermion matrix which quickly increase the
numerical cost. A detailed description of the terms used in our $D$ and the
values of the coefficients are given in the appendix in  Table \ref{terms} and
Table \ref{coeffvals}. Besides the three boundary  conditions (\ref{rendbc}),
(\ref{improve}) we implement 14 equations from  the system (\ref{qsyst}) to
match the number of expansion coefficients (17). These equations correspond to
the shortest terms with paths on the hypercube contributing to the expansion
(\ref{eexp}) of $E$. We remark, that it was  straightforward to exactly solve
the equations (\ref{qsyst}), (\ref{rendbc}),  (\ref{improve}) with a standard
solver \cite{numrec} and we also found that the system is very stable, i.e.~we
found only a single solution for our setting.

\setcounter{equation}{0}
\section{Properties of our Dirac-operator}

In this section we analyze the properties of our Dirac operator for different
ensembles of background gauge fields. We start with the analysis of the free 
case, followed by a study using quenched SU(3) configurations generated with
the standard  Wilson action for the gauge fields. We conclude with analyzing
gauge fields from the improved L\"uscher-Weisz action.

\subsection{The free case}

\begin{figure}[ht]
\begin{center}
\epsfysize=9cm \epsfbox[ 51 195 492 630 ] {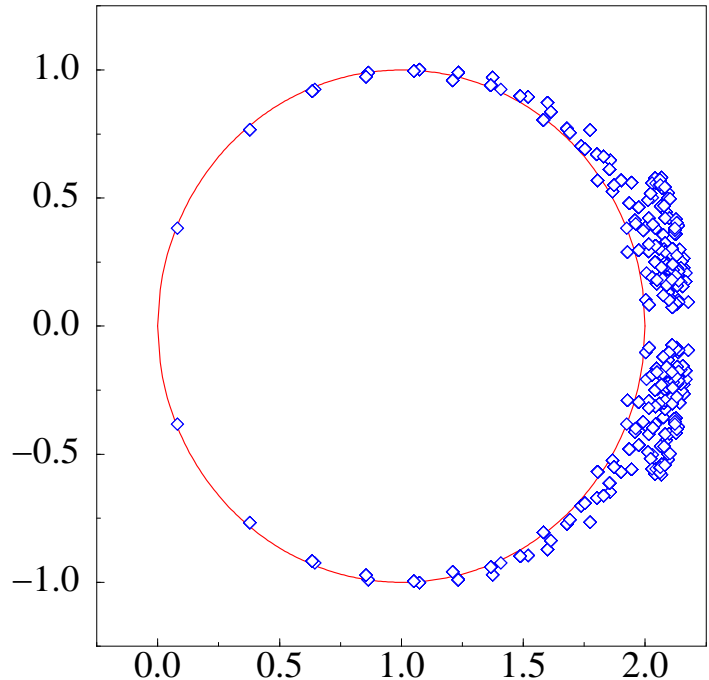}
\caption{Spectrum of our Dirac operator for the free case in
the complex plane. The represent the eigenvalues. We also show the 
Ginsparg-Wilson circle, i.e.~the circle of radius 1 and center 1.
\label{freespec}}
\end{center}
\end{figure}

For the free case it is possible to completely diagonalize the Dirac  operator
using Fourier transformation. In Fig.~\ref{freespec} we show the free spectrum
of our $D$ in the complex plane. In addition  to the eigenvalues (symbols) we
also show the Ginsparg-Wilson circle  (full curve)  which supports the spectrum
of any exact solution of (\ref{giwi}).  Throughout this article we present our
spectra for $8^4$ lattices. This is a relatively small volume for usual
standards, however, the Dirac operator studied here is ultralocal and its 
extent is well within the lattice size.

Let us briefly discuss some features of the free spectrum. It is obvious, that 
our approach optimizes the alignment of the eigenvalues along the circle for
the eigenvalues in the physical branch, i.e.~the eigenvalues in the  vicinity
of the origin.  The alignment of the eigenvalues in the doubler branches is
less perfect. The important feature, however, is the clear separation of the
doubler modes and the physical eigenvalues.   We furthermore found, that adding
additional terms in (\ref{dexp})  systematically improves the situation also
for the doublers.  This property of our approach of first optimizing the
physical branch and aligning the doubler modes along  the circle when adding
higher terms persists also when analyzing the  spectrum for non-trivial
background gauge fields and was observed also for  the 2-d case
\cite{GaHi00}.   

At this point it is interesting to compare our method to the study of 
truncated perfect actions for free fermions  \cite{Biholz1,fpa,Biholz2}. For
free  fermions the perfect action can be computed explicitly \cite{fixpd1}  as
an infinite series and is a solution of the  Ginsparg-Wilson equation. After
truncation to e.g.~only terms on the  hypercube the result is an approximate
solution of the Ginsparg-Wilson equation for the free case. 

This truncated perfect action will however differ  from our approach in two
important aspects: Firstly actions for  free fermions only contain scalar and
vector terms. Terms in  the tensor, pseudovector and pseudoscalar  sectors
vanish identically when no gauge field is coupled.  Thus e.g.~terms in the
tensor sector which are necessary for ${\cal O}(a)$ improvement and also play 
an important role for a smooth spectrum in the physical branch  \cite{GaHi}
have to be included a posteriori. Secondly for each  possible endpoint of a
fermion path  the truncated perfect action for free fermions provides only a
single coefficient. On the other hand, our symmetry analysis (compare
(\ref{dexp})) shows, that different paths leading to this endpoint can come
with  different coefficients. Thus the distribution of the single coefficient
of the free perfect action among the various coefficients allowed by the exact
symmetry analysis is an additional task. 

Overcoming these problems, Bietenholz \cite{Biholz2} presented  interesting
results for an approximate solution of the Ginsparg-Wilson  equation in 4-d.
After introducing fat links and a link amplification factor for the
coefficients of the paths, an approximate solution of the Ginsparg-Wilson
equation also for the case with  gauge fields is obtained from the truncated
perfect action.  It is interesting  to note that the spectra presented in
\cite{Biholz2} show a behavior quite contrary to the spectra for our $D$: They
come with a  very smooth doubler branch while the modes in the physical branch
have quite  large fluctuations. Bietenholz expresses confidence that  the
physical branch can be smoothened by adding tensor terms but expects that these
terms will at  the same time destroy the good alignment of the eigenvalues in
the doubler branch \cite{Biholz2}.

\subsection{Results for quenched gauge configurations generated 
with the Wilson gauge action}

After having studied the properties of our $D$ for the free case we are  now
analyzing the Dirac operator in ensembles of quenched SU(3) gauge field 
configurations.  We concentrate on the branch of the spectrum near small
eigenvalues, which is the one most relevant in the continuum limit. We compute
the eigenvalues in  this physical  branch of the spectrum and compare  them
with the eigenvalues  for the Wilson operator (\ref{dwil}). The values of
$\kappa$ used for the Wilson operator (\ref{dwil}) can be found in
Table~\ref{rundat} and will be discussed below.

\begin{figure}[t]
\begin{center}
\epsfysize=8.7cm \epsfbox[ 11 71 454 649 ] {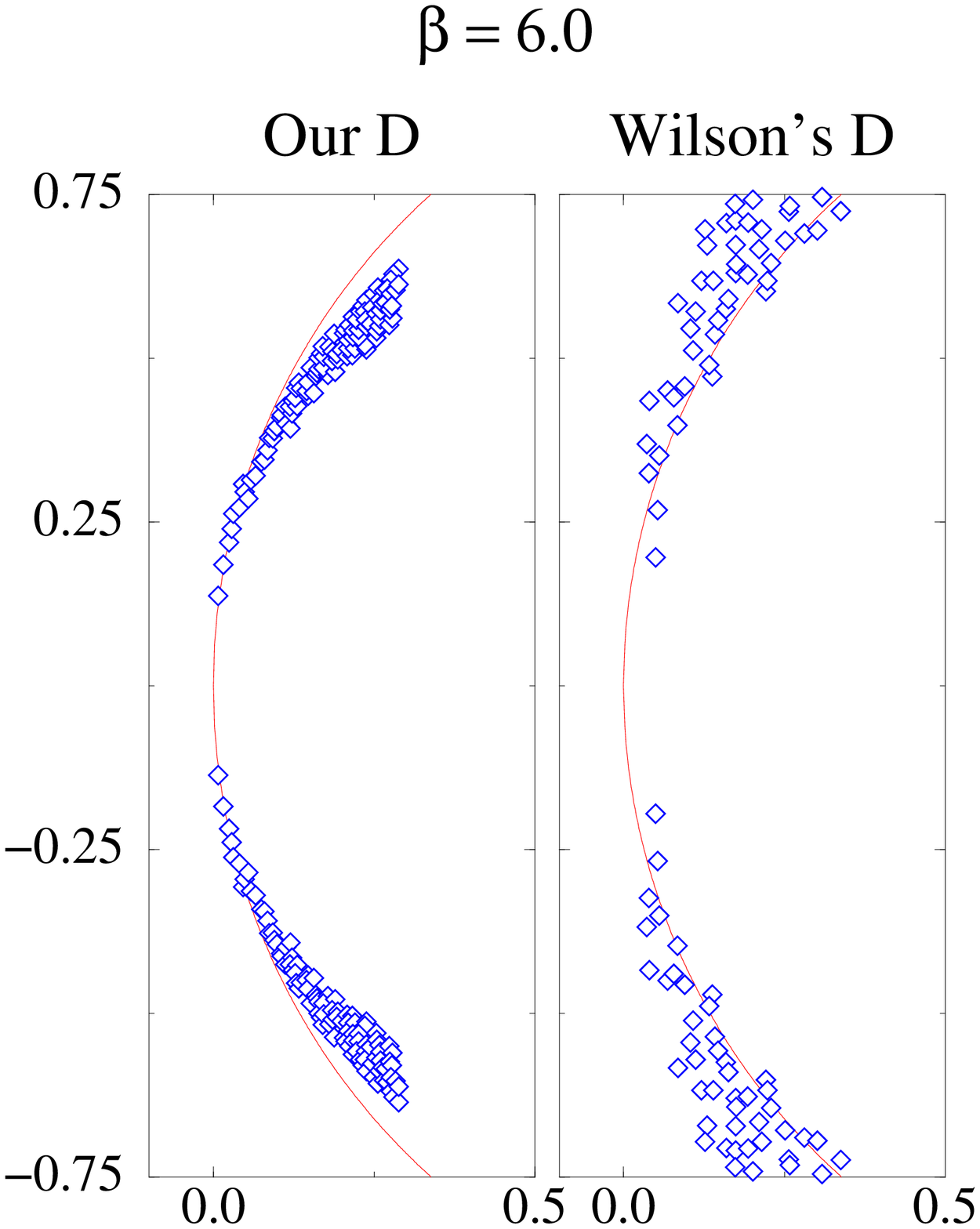}
\epsfysize=8.7cm \epsfbox[ 73 71 454 649 ] {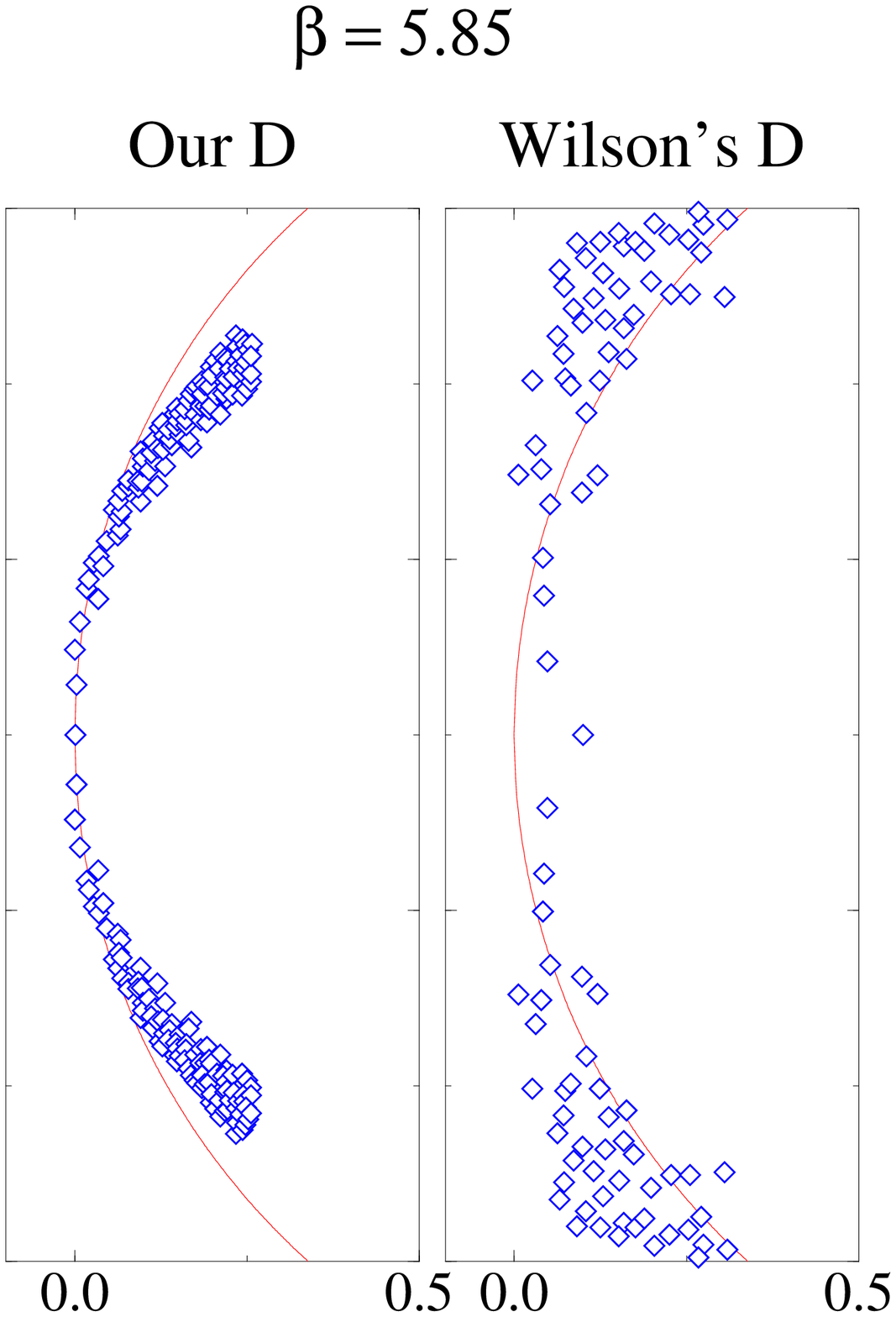}
\caption{The physical branch of spectra of Dirac operators in
the complex plane for quenched SU(3) gauge field configurations generated
using the Wilson gauge action. From left to right we show:
The spectrum of our $D$ at $\beta = 6.0$, for Wilson's
Dirac operator at $\beta = 6.0$, our $D$ at $\beta = 5.85$ and finally 
for Wilson's $D$ at $\beta = 5.85$.
\label{spectraw}}
\end{center}
\end{figure}

For the case of non-trivial background gauge fields the Dirac operator can no
longer be diagonalized completely unless one uses very small lattices.  Here we
use the Implicitly Restarted Arnoldi Method  \cite{arnoldi} to compute
eigenvalues in the physical branch of the spectrum, i.e.~near the origin. The
method allows to specify a search criterion for the eigenvalues and we use this
feature to compute the eigenvalues with the smallest real parts. In order to
get  insight into the typical behavior of the spectrum  of our $D$, we analyzed
the eigenvalues on 4 ensembles of quenched gauge fields. Each ensemble consists
of 20  well decorrelated configurations. Two of the ensembles were generated
with  the standard Wilson action at $\beta = 6.0$ and $\beta = 5.85$. They will
be discussed in this subsection. The other two ensembles were generated using
the  improved L\"uscher-Weisz action \cite{LuWeact} and will be analyzed in
the  next subsection.   For each gauge field configuration we computed  60
eigenvalues in the physical branch for both our Dirac operator as well as for
the Wilson-Dirac operator. For the plots in Fig.~\ref{spectraw} and
Fig.~\ref{spectralw} we increased the number of computed  eigenvalues to 200.
All these  calculations were done on $8^4$ lattices with anti-periodic
boundary  conditions for the Dirac operator.  

In Fig.~\ref{spectraw} we show the physical branch of the spectrum of our
operator for  quenched gauge field configurations generated with the standard
Wilson gauge action at  $\beta = 6.0$  and $\beta = 5.85$. We compare these
spectra with the spectra of the Wilson-Dirac operator on the same
configurations. The symbols are the  numerically computed eigenvalues and the
full curve is the Ginsparg-Wilson circle. 

It is obvious, that for our $D$ the eigenvalues are more ordered 
than for Wilson's $D$ for both $\beta = 6.0$ and $\beta = 5.85$. The 
spread of the eigenvalues of our $D$ is reduced when compared to the 
Wilson case and the eigenvalues tend to order near a curve. 
However, what is also 
obvious is the fact that although this curve touches the origin, it also bends
away from the Ginsparg-Wilson circle as the size of the imaginary parts
increases. As can be seen from  the Fourier transform in the free case, it is
essentially the   vector terms in the expansion (\ref{dexp}) contributing to  
the imaginary parts of the eigenvalues. As we have already remarked above, we
allowed only for a relatively small number of vector terms  (compare Table
\ref{terms}) in order to reduce the numerical cost. We thus attribute the
bending away of the  eigenvalues to this reduced approximation of the vector
sector. In the next  section we will, however, demonstrate, that our
parametrization of the vector sector is sufficient if one uses improved gauge
field actions.

It is also interesting to study the behavior of the real eigenmodes -- the
$\beta = 5.85$ part of Fig.~\ref{spectraw} shows one real eigenvalue. 
Typically the fluctuations of the real eigenvalues of the Wilson-Dirac operator
are relatively large. In particular for so-called exceptional configurations 
(in the sense of a breakdown of the matrix inversion) a
real mode has fluctuated so heavily towards small values  that it compensates
for the bare quark mass and the resulting zero mode leads to a breakdown of the
matrix inversion. We found, and Fig.~\ref{spectraw} shows an example, that our
$D$ strongly suppresses the fluctuations  of the real modes and turns them into
almost perfect zero modes. We thus expect that at nonzero bare quark mass our
$D$ considerably  reduces the numerical difficulties with exceptional
configurations. 

Let us now discuss the overall fluctuation of the rest of the physical branch
of the spectrum. An exact solution of the Ginsparg-Wilson equation has its
spectrum on the Ginsparg-Wilson circle and only the density of eigenvalues on
this circle will fluctuate. The eigenvalues of  the Wilson-Dirac operator, on
the other hand, have relatively large fluctuations in all directions.  When
analyzing the ensemble of 20 configurations for both  $\beta = 6.0$ and $\beta
= 5.85$ we found that for our $D$ the fluctuations of the physical edge of the 
spectrum are considerably suppressed when compared to the Wilson-Dirac 
operator. In order to quantify this statement we fit a circle of the  form 
\begin{equation} 
y \; = \; \pm \; i \; \sqrt{ 1 \; - \; ( \; 1 \; + \;
\varepsilon  \; - \; x \; )^2} \; , 
\label{circle} 
\end{equation} 
to the 10 smallest eigenvalues in the physical branch omitting the exactly real
eigenvalues, because they deviate significantly in the case of the Wilson
operator (compare above). Eq.~(\ref{circle}) describes a circle of radius 1
around $(1+\varepsilon,0)$ in the complex plane.  The parameter $\varepsilon$
of the fit gives the position where  the circle crosses the real axis. The
fluctuation of $\varepsilon$  provides a measure for the fluctuation of the
eigenvalues in the physical branch of the spectrum. 

\begin{figure}[ht]
\begin{center}
\epsfig{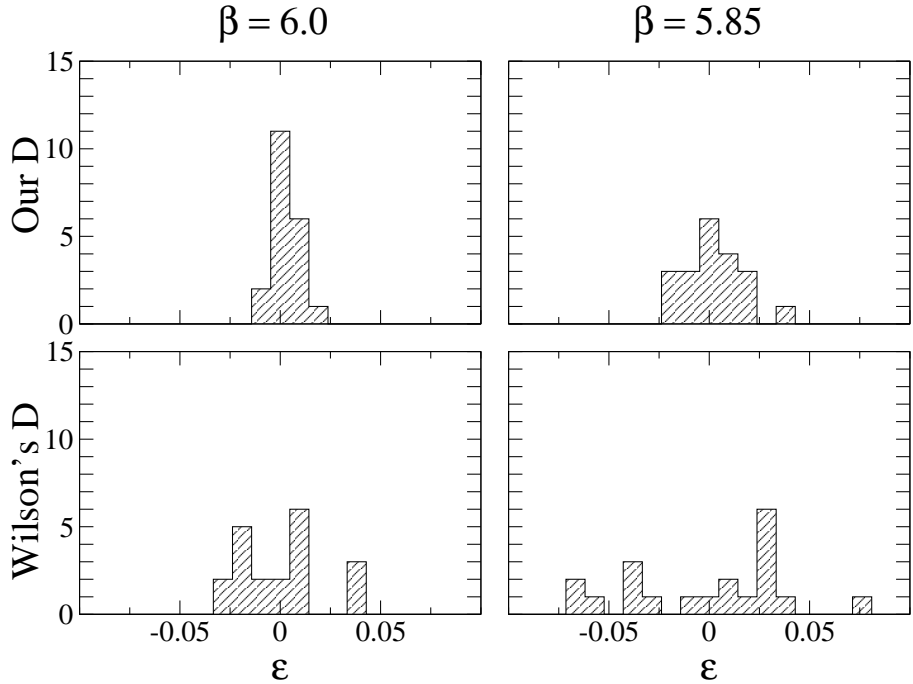}
\caption{Histograms indicating the fluctuation of $\varepsilon$ (i.e.~the
fluctuation of the physical branch of the spectrum) for two ensembles  of 20
quenched gauge field configurations  generated with the Wilson gauge action at
$\beta = 6.0$  (left column) and $\beta = 5.85$ (right column). The top row
shows the results for our $D$, the bottom row displays the results for the 
Wilson-Dirac operator.
\label{physedgew}}
\end{center}
\end{figure}

In Fig.~\ref{physedgew} we show for each of our ensembles a histogram of the
values of $\varepsilon$ for all 20 configurations, comparing the results for
our Dirac operator  with those for Wilson's $D$.  It can be seen, that our $D$
has considerably reduced the fluctuations of the physical edge of the spectrum,
in particular for $\beta = 6.0$.

At this point we can also comment on the determination of the critical $\kappa$
for the Wilson-Dirac operator (\ref{dwil}). Since the first term is 
proportional to the unit matrix, it is sufficient to diagonalize the  hopping
matrix $H$. The value of $\kappa$ was then adjusted such that the distribution
of $\varepsilon$ is centered at 0. The values  of $\kappa$ obtained for our 4
ensembles are given in Table \ref{rundat} below.

\subsection{The effect of improving the gauge action}

In this section we discuss the effect of improving the gauge field on the
spectral properties of the Dirac operator. In the context of  domain wall
fermions it has been found that improving the gauge field action allows to work
with a smaller extension of the 5-th direction \cite{gaugeimp}. Also for the
4-d setting it is worth testing if  improving the gauge field action leads to
an improvement of the spectral  properties of $D$ in particular since from a
numerical point of view improving the gauge action is an inexpensive  measure.

\begin{table}[t]
\begin{center}
\begin{tabular}{cccc}
\hline
 & \hspace{1mm} free case \hspace{1mm}
& Wilson action & L\"uscher-Weisz action \\
 & & $\beta = 6.0$ \hspace{4.5mm} $\beta = 5.85$ & 
$\beta_1 = 8.45$ \hspace{4.5mm} $\beta_1 = 8.15$ \\
\hline
 $ \frac{1}{3} <U_{pl}>$   & 1.0 & 0.594(1) \hspace{5mm} 0.576(1) &
 0.652(1) \hspace{5mm} 0.633(1)    \\
 $ \kappa $    & 0.5 & 0.624 \hspace{10mm} 0.642 & 0.607 \hspace{10mm} 
0.623  \\
\hline
\end{tabular}
\end{center}
\caption{Parameters for our ensembles of quenched gauge field configurations
and the free case. We list the
expectation value $<U_{pl}>/3$ of the plaquette and the critical  $\kappa$ 
we use for the Wilson operator.
\label{rundat}}
\end{table}

In order to analyze the effects of gauge improvement we generated 2 ensembles
of 20 gauge configurations each, using the L\"uscher-Weisz  action
\cite{LuWeact,Aletal95}.  The parameters were adjusted such that the  physical
scale (the effective 
string tension) approximates the two scales of our ensembles from
the standard Wilson gauge action at $\beta = 6.0$ and $\beta = 5.85$. Our 
estimator for the effective
string tension is based on Wilson loops up to extent 3 only
and therefore not to be compared with the values derived on larger 
lattices with a detailed finite size analysis. However, our principal 
incentive was not to obtain such a precise value, but to identify the values 
of the gauge couplings where the two gauge actions have roughly the same
scale.

To be specific, we use the setting for the improved gauge action as
presented in \cite{Aletal95} but with only the rectangle term. Explicitly, the 
gauge field action reads
\begin{equation}
S[U] \; = \; \beta_1 \sum_{pl} \frac{1}{3} \mbox{Re~Tr} ( 1 - U_{pl} ) 
\; + \; 
\beta_2 \sum_{rt} \frac{1}{3} \mbox{Re~Tr} ( 1 - U_{rt} ) \; ,
\label{sgauge}
\end{equation}
where the first sum is over all plaquettes and the second sum over all  $2
\times 1$ rectangles. $\beta_1$ is the driving parameter while  $\beta_2$ can
be computed from $\beta_1$  using tadpole improved perturbation theory 
\cite{LeMa93} giving \cite{Aletal95} 
\begin{equation}
\beta_2 \; = \; - \; \frac{ \beta_1}{ 20 \; u_0^2} \; 
[ 1 + 0.4805 \, \alpha_s ]
\; .
\end{equation}
with
\begin{equation}
u_0 \; = \; \Big( \frac{1}{3} \mbox{Re~Tr} \langle U_{pl} \rangle 
\Big)^{1/4} \; \; , \; \; \alpha_s \; = \; - \;
\frac{ \ln \Big(\frac{1}{3} \mbox{Re~Tr} \langle U_{pl} \rangle 
\Big)}{3.06839} \; .
\end{equation}
The coupling $\beta_1$ is determined self-consistently with $u_0$  and
$\alpha_s$ for a given $\beta_1$.  We adjust the parameter $\beta_1$ such
that the  physical scale approximates the two scales of our ensembles from the
standard Wilson gauge action at $\beta = 6.0$ and $\beta = 5.85$, i.e. such
that the effective string tension of the improved ensemble (as discussed above)
roughly matches the value of
the ensemble from the standard Wilson action. The resulting values are 
$\beta_1 = 8.45$ which corresponds to $\beta = 6.0$ for the Wilson  action and
$\beta_1 = 8.15$ corresponding to $\beta = 5.85$. We list our results for the
pla\-quette expectation value
and the values of the  critical $\kappa$ determined for the Wilson-Dirac
operator in  Table \ref{rundat}.

\begin{figure}[t]
\begin{center}
\epsfysize=8.7cm \epsfbox[ 11 71 454 649 ] {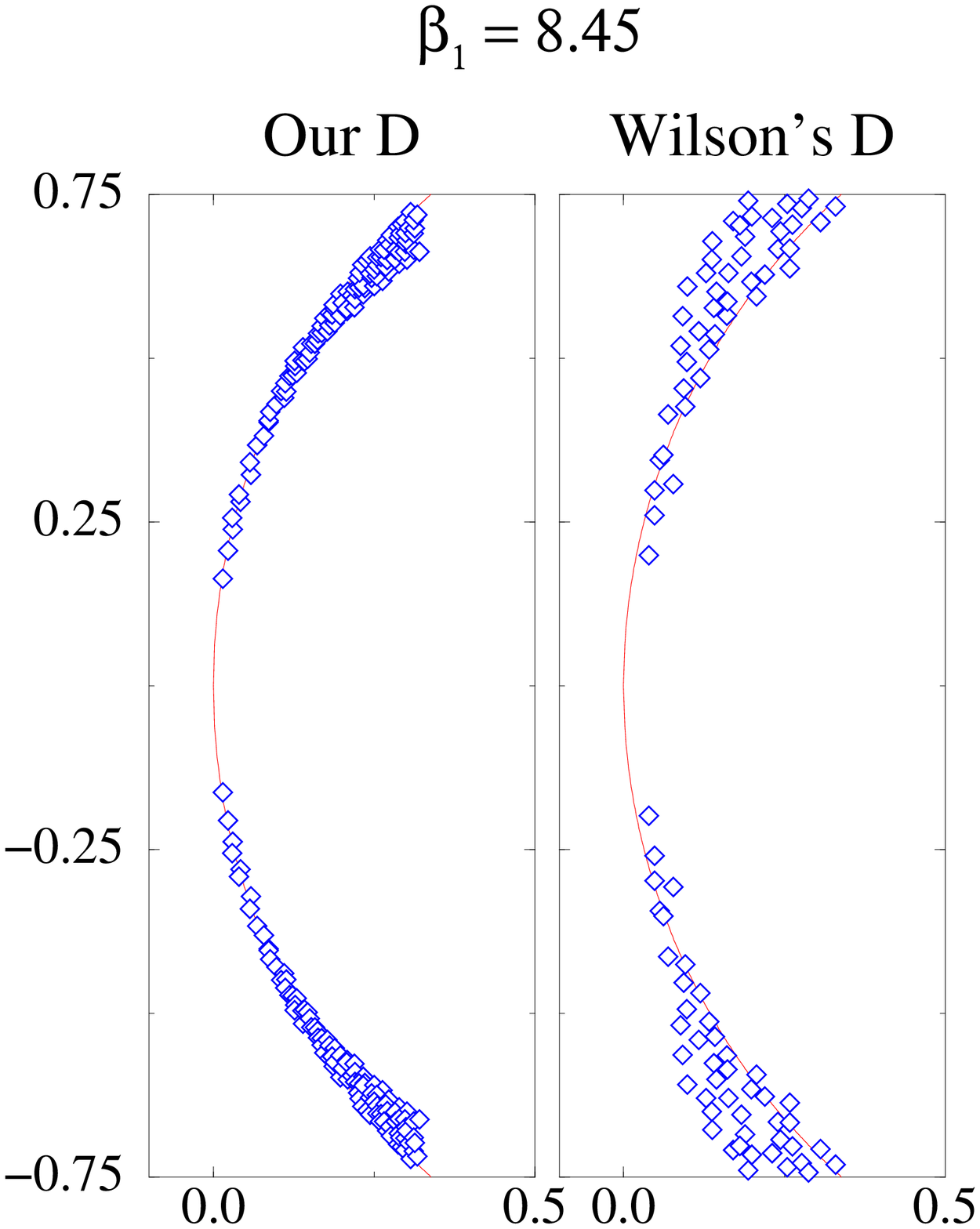}
\epsfysize=8.7cm \epsfbox[ 73 71 454 649 ] {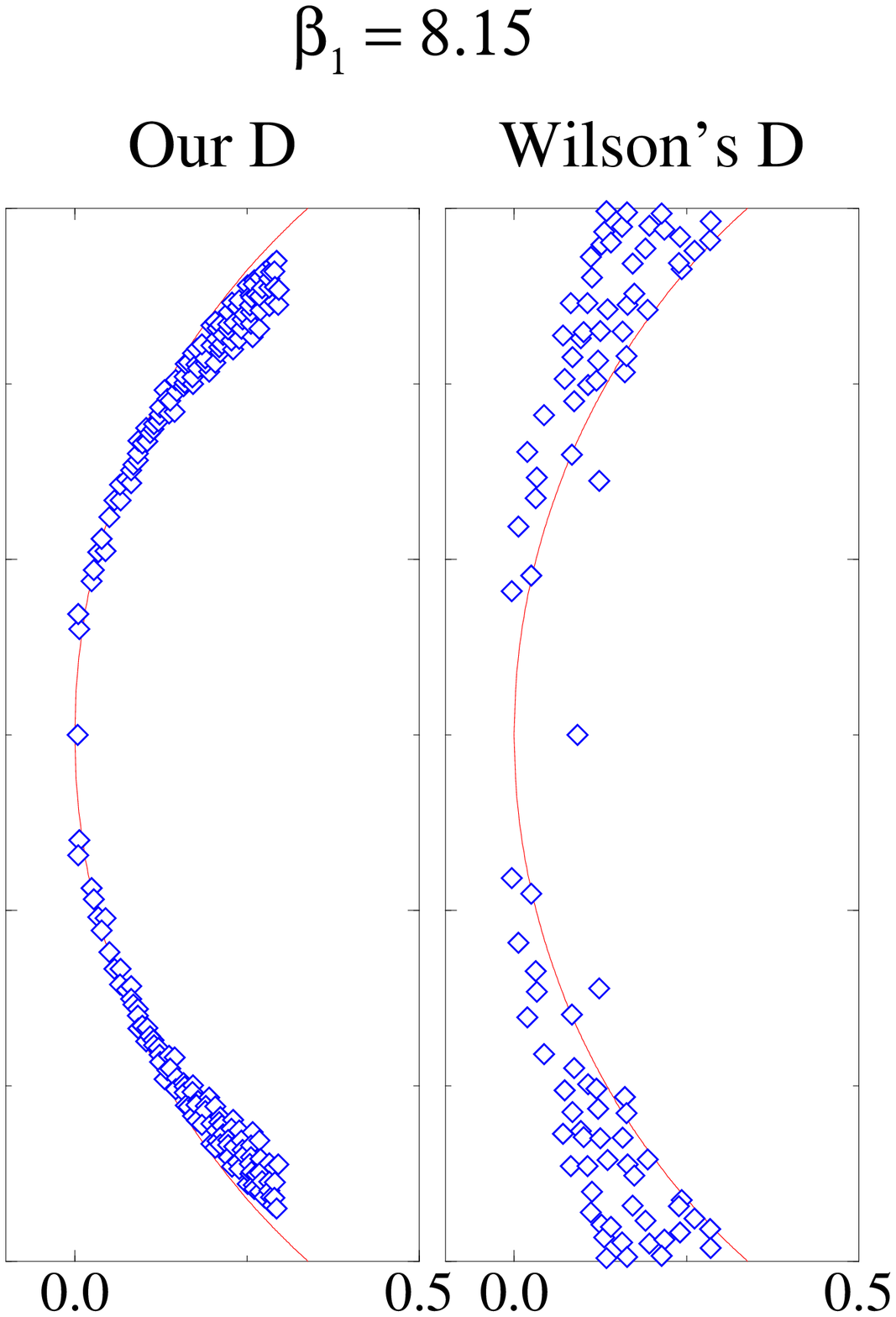}
\caption{The physical branch of spectra of Dirac operators in the complex plane
for quenched SU(3) gauge field configurations generated using the
L\"uscher-Weisz gauge action. From left to right we show: The spectrum of our
$D$ at $\beta_1 = 8.45$, for Wilson's Dirac operator at $\beta_1 = 8.45$, our
$D$ at $\beta_1 = 8.15$ and finally  again Wilson's $D$ at $\beta_1 = 8.15$.
\label{spectralw}
}
\end{center}
\end{figure}

Like for the ensembles generated with the Wilson gauge action also here we
computed for each configuration 60 eigenvalues in the physical branch for both,
our $D$  as well as for Wilson's Dirac operator. When inspecting these 
eigenvalues, one finds that the spectral properties of our $D$ are 
considerably improved. In Fig.~\ref{spectralw} we show examples  of spectra for
one configuration from each of the ensembles.  As in the last section we show
200 eigenvalues in the physical branch for our $D$ as well as for the
Wilson-Dirac operator both for the same gauge configuration. 

\begin{figure}[t]
\begin{center}
\epsfig{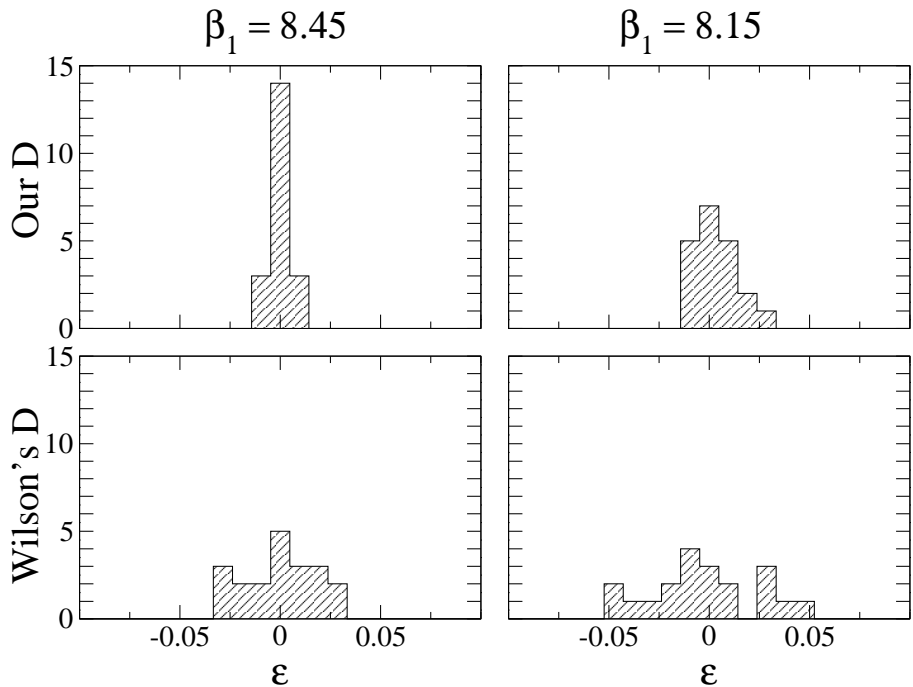}
\caption{Fluctuation  of the physical branch of the spectrum for the two
ensembles  from the L\"uscher-Weisz action at $\beta_1 = 8.45$  (left column)
and $\beta_1 = 8.15$ (right column). The top row shows the results for our $D$,
the bottom row displays the results for the  Wilson-Dirac operator.
\label{physedgelw}}
\end{center}
\end{figure}

When comparing these plots to Fig.~\ref{spectraw} one finds that the 
eigenvalues of our $D$ are somewhat better 
aligned along a single curve, i.e.~the
distribution is narrower.  Also the imaginary parts are larger and the curve is
following the Ginsparg-Wilson circle much closer than it was the case for the 
standard Wilson gauge action.  The improved spectral properties of $D$   can be
attributed to the following feature of the improved gauge field ensemble: When
comparing the plaquette expectation value of the improved ensemble with that of
the Wilson ensemble at the same physical  scale one finds that improvement
brings the plaquettes considerably closer  to 1 (see Table \ref{rundat}).  A
second, although smaller effect, might be the better rotational properties of
the gauge fields from the improved action \cite{Aletal95}. 

We remark, that also the eigenvalues of the  Wilson-Dirac operator are slightly
more ordered when using the improved gauge field action. It is expected that 
this already could help to improve on the problems with the  singularities in
the inverse square root of the overlap operator  \cite{liu}. 

We conclude this subsection with repeating last section's analysis of the 
fluctuations of the physical branch, now for the improved ensemble.  When
comparing Fig.~\ref{physedgelw} and Fig.~\ref{physedgew} one  finds, that for
both our $D$ as well as for the Wilson-Dirac operator  the distribution of
$\varepsilon$ becomes slightly narrower when using the improved gauge action. 

We thus find that using the improved gauge action is a numerically inexpensive
measure leading to a considerable improvement of the spectral  properties of
the lattice Dirac operator. In particular we observe that  improving the gauge
fields allows to work  with a relatively small number of terms in the vector
sector of $D$ and still obtain a good approximation of a solution of the
Ginsparg-Wilson  equation. 

\setcounter{equation}{0}
\section{Discussion}

In this article we have presented first tests in 4-d for a new method of 
constructing approximate solutions of the Ginsparg-Wilson equation. The  most
general Dirac operator on the lattice is systematically expanded in a  series
of simple operators. The Ginsparg-Wilson equation turns into a set of  coupled
quadratic equations for the expansion coefficients. For a finite 
parametrization of $D$ these equations can be solved and the corresponding  $D$
is an approximate solution of the Ginsparg-Wilson equation. We implement
boundary conditions which allow to work with a very economical parametrization
of $D$ and still obtain a good approximation of a solution of the 
Ginsparg-Wilson equation. Our framework allows to systematically include 
${\cal O}(a)$ improvement.

We find that our method has the intriguing feature that already a few terms
lead to a good alignment on the Ginsparg-Wilson circle for the physical  modes.
The alignment of the doubler branch can be systematically improved  by adding
additional terms in the parametrization of $D$. We have demonstrated that the
fluctuations of the eigenvalues in the physical branch are much smaller than
e.g.~for the Wilson-Dirac operator and we expect that our $D$ gets  essentially
rid of the problems with exceptional configurations. Finally we have observed
that using an improved action for the gauge fields  considerably improves the
spectral properties of our $D$.

At this point we would like to comment on the numerical cost of our Dirac
operator. So far we have not implemented an optimized  matrix-vector
multiplication for our $D$. Our test program was kept general in order to be
able to test Dirac operators with more terms than the ones we present here.
Thus we give a theoretical estimate of the numerical cost of our Dirac operator
and compare it to the cost of the standard Wilson-Dirac operator. To leading
order the numerical cost is determined by the number of non-vanishing entries
in the fermion matrix. For our $D$ this number is given by 297 where we have
factored out the volume and a factor of 12 for color and spinor structure. The
corresponding number for Wilson's Dirac operator is 17. Thus to leading order
we find a factor of 297/17 = 17.47 when comparing the numerical cost for our
$D$ with Wilson's Dirac operator. This estimate, however, assumes that the all
necessary products of the link variables can  be stored. This is possible for
smaller lattices up to $10^4$, but for larger  lattices only pieces of paths
can be stored. With a good storage strategy the additional cost for building up
the longer paths can be kept below a factor of 2. We  estimate our $D$ to be
approximately 25 times more expensive when compared to Wilson's $D$. Finally,
we also found that  the Arnoldi diagonalization  routine we use typically needs
35 \% less matrix multiplications for our $D$  as compared to Wilson's Dirac
operator.

Let us now come back to the three possible applications of our $D$ which we 
mentioned in the introduction:
\begin{itemize}
\item We think that it would be interesting to use our $D$ for analyzing the
pion spectrum in the quenched approximation. The fact that the fluctuations of
the spectrum near the origin are highly suppressed should allow one  to perform
simulations at small bare quark masses without running into  problems with
exceptional configurations. Also studies of the distribution 
of the eigenvalues  of our $D$ could be
used to compute the chiral condensate from the Banks-Casher formula. 
\item  Our successful construction of an approximate solution for the
Ginsparg-Wilson equation also seems to be good news for the project of
constructing perfect fermion actions using block spin transformations. It seems
quite feasible to construct a sufficiently rich parametrization with  only 15
to 20 parameters. It seems to us that in the scalar sector it is necessary to
include relatively many terms, while in the higher sectors one can try to be
more economical, in particular when using improved or perfect gauge actions.
\item  Finally our $D$ is also a good candidate for an  improved starting
operator $D_0$ in the overlap projection (\ref{overlapd}): It is already  much
closer to a solution of the Ginsparg-Wilson equation than e.g.~the  standard
Wilson-Dirac operator which is usually used for $D_0$. Thus we expect that an
expansion for the inverse square root converges faster  when using our $D$ as
starting operator.
\end{itemize}

\vskip5mm
\noindent
{\bf Acknowledgements: } The authors would like to thank Mark Alford, Wolfgang
Bietenholz, Richard Brower,  Peter Hasenfratz, Keh-Fei Liu, Ferenc Niedermayer,
Jack Verbaarschot and Uwe-Jens Wiese for interesting discussions. We also thank
Mark Alford and Peter Lepage for sharing their computer program for  generating
the improved gauge field ensembles.  The numerical computations were done on
the NICse Alpha Linux cluster  at NIC J\"ulich. Christof Gattringer thanks the
INT at  the University of Washington in Seattle, where part of this work was
done,  for its kind hospitality.

\newpage
\begin{appendix}
\setcounter{equation}{0}
\section{Technical appendix}
In this appendix we describe in more detail the terms in our Dirac operator 
and give the values for the coefficients which we use for the 4 ensembles of
quenched gauge field configurations. 

As has been pointed out in  Section 2.1, the most general Dirac operator $D$
can be expanded in the  series (\ref{dexp}). Each term in this series is
characterized by three  pieces:
\begin{enumerate}
\item a generator of the 
Clifford algebra,
\item a group of paths, 
\item a real coefficient. 
\end{enumerate}
The paths within a group can have different signs which are determined by the
symmetries, C, P,  $\gamma_5$-hermiticity and rotation invariance (for the
operation of these  symmetries see Section 2.1). The symmetries also determine
which paths are grouped together. Thus it is sufficient to characterize a group
of paths by a single generating path and all the other paths in the group as
well as their relative their sign factors can be  determined by applying the
symmetries. 

In addition, for the vector and tensor terms appearing in our $D$ it is 
sufficient to give the paths only for one vector (tensor) since rotation
invariance immediately fixes the structure for the other vector (tensor) 
terms. In order to describe our $D$ we start with listing the three determining
pieces for each term in Table~\ref{terms}. 

\begin{table}[p]
\begin{center}
\begin{tabular}{ c c c }
\hline
 & & \\
Clifford generator & Generating path & Name of coefficient \\
 & & \\
\hline
1\hspace{-1.0mm}I & $ < > $ & $s_1$ \\
1\hspace{-1.0mm}I & $ <1> $ & $s_2$ \\
1\hspace{-1.0mm}I & $ <1,2 > $ & $s_3$ \\
1\hspace{-1.0mm}I & $ <1,2,3 > $ & $s_5$ \\
1\hspace{-1.0mm}I & $ <1,1,2 > $ & $s_6$ \\
1\hspace{-1.0mm}I & $ <1,2,-1 > $ & $s_8$ \\
1\hspace{-1.0mm}I & $ <1,2,3,4 > $ & $s_{10}$ \\
1\hspace{-1.0mm}I & $ <1,2,-1,3 > $ & $s_{11}$ \\
1\hspace{-1.0mm}I & $ <1,2,-1,-2 > $ & $s_{13}$ \\
\hline
$ \gamma_1$ & $ <1> $ & $v_1$ \\
$ \gamma_1$ & $ <1,2> $ & $v_2$ \\
$ \gamma_1$ & $ <1,2,3> $ & $v_4$ \\
$ \gamma_1$ & $ <2,1,3> $ & $v_5$ \\
\hline
$ \gamma_1 \gamma_2 $ & $ <1,2> $ & $t_1$ \\
$ \gamma_1 \gamma_2 $ & $ <1,2,3> $ & $t_2$ \\
$ \gamma_1 \gamma_2 $ & $ <1,3,2> $ & $t_3$ \\
$ \gamma_1 \gamma_2 $ & $ <1,2,-1> $ & $t_5$ \\
\hline
\end{tabular}
\end{center}
\caption{Description of the terms in our $D$.}
\label{terms}
\end{table}

In Section 4 we analyze our $D$ in different ensembles of  quenched background
gauge fields. In particular  we use ensembles generated with the standard
Wilson action at $\beta = 6.0$ and $\beta = 5.85$ and ensembles from the
L\"uscher-Weisz action at $\beta_1 = 8.45$ and $\beta_1 = 8.15$ (for more
details see Section 4.3). For all these configurations we use the same
parametrization of $D$. Only the values of the coefficients $s_i^\alpha,
v_i^\alpha$ and $t_i^\alpha$ and the two  normalization factors $z_s$ and $z_v$
for the boundary conditions  differ. In Table~\ref{coeffvals} we list their
values. 

\begin{table}[hp]
\begin{center}
\begin{tabular}{ c c c c }
\hline
 & \hspace{2mm} free case \hspace{2mm}
& Wilson action & L\"uscher-Weisz action \\
 & & $\beta = 6.0$ \hspace{5.5mm} $\beta = 5.85$ & 
$\beta_1 = 8.45$ \hspace{5.5mm} $\beta_1 = 8.15$ \\
\hline
 $s_1$  & $+1.488513$ & $+1.541745$ \hspace{2mm} $+1.536335$ & 
$+1.545142$ \hspace{2mm} $+1.546343$   \\
 $s_2$  & $-0.030753$ & $-0.066240$ \hspace{2mm} $-0.069085$ &
$-0.061723$ \hspace{2mm} $-0.063831$   \\
 $s_3$  & $-0.011132$ & $-0.013808$ \hspace{2mm} $-0.014036$ &
$-0.014045$ \hspace{2mm} $-0.013612$   \\
 $s_5$  & $-0.002128$ & $-0.002763$ \hspace{2mm} $-0.002844$ &
$-0.002571$ \hspace{2mm} $-0.002690$   \\
 $s_6$  & $-0.000691$ & $+0.002396$ \hspace{2mm} $+0.002649$ &
$+0.002228$ \hspace{2mm} $+0.002198$   \\
 $s_8$  & $-0.005842$ & $-0.005813$ \hspace{2mm} $-0.005786$ &
$-0.005415$ \hspace{2mm} $-0.005811$   \\
 $s_{10}$ & $-0.000303$ & $-0.000557$ \hspace{2mm} $-0.000593$ & 
$-0.000512$ \hspace{2mm} $-0.000526$   \\
 $s_{11}$ & $-0.000830$ & $-0.001217$ \hspace{2mm} $-0.001241$ &
$-0.001181$ \hspace{2mm} $-0.001200$   \\
 $s_{13}$ & $-0.010061$ & $+0.007853$ \hspace{2mm} $+0.007918$ &
$+0.007831$ \hspace{2mm} $+0.007788$    \\
\hline
 $v_1$  & $+0.032416$ & $+0.091060$ \hspace{2mm} $+0.096935$ & 
$+0.107048$ \hspace{2mm} $+0.087543$   \\
 $v_2$  & $+0.032158$ & $+0.022333$ \hspace{2mm} $+0.022054$ & 
$+0.017838$ \hspace{2mm} $+0.022207$   \\
 $v_4$  & $+0.008365$ & $+0.006828$ \hspace{2mm} $+0.006007$ & 
$+0.007926$ \hspace{2mm} $+0.007540$  \\
 $v_5$  & $-0.013326$ & $+0.000299$ \hspace{2mm} $+0.000423$ & 
$+0.001779$ \hspace{2mm} $-0.000364$  \\
\hline
 $t_1$  & $-0.108727$ & $-0.103545$ \hspace{2mm} $-0.103318$ & 
$-0.102314$ \hspace{2mm} $-0.103576$   \\
 $t_2$  & $-0.002129$ & $-0.003544$ \hspace{2mm} $-0.003715$ & 
$-0.003206$ \hspace{2mm} $-0.003394$   \\
 $t_3$  & $+0.001356$ & $+0.002333$ \hspace{2mm} $+0.002444$ & 
$-0.002198$ \hspace{2mm} $+0.002239$   \\
 $t_5$  & $-0.006355$ & $-0.006945$ \hspace{2mm} $-0.007266$ & 
$-0.006170$ \hspace{2mm} $-0.006645$   \\
\hline  
$z_s$ & 1.0 & 0.876 \hspace{10mm} 0.866 & 0.895 \hspace{10mm} 0.885 \\
$z_v$ & 1.0 & 0.868 \hspace{10mm} 0.885 & 0.846 \hspace{10mm} 0.854 \\
\hline
\end{tabular}
\end{center}
\caption{The numerical values of the 
coefficients $s_i^\alpha, v_i^\alpha$ and $t_i^\alpha$ and the
factors $z_s$ and $z_v$ for
the different ensembles of gauge fields.
\label{coeffvals}}
\end{table}

\end{appendix}

\end{document}